\documentclass[%
 reprint,
superscriptaddress,
showpacs,
 amsmath,amssymb,
 aps,
]{revtex4-1}

\usepackage{graphicx}
\usepackage{dcolumn}
\usepackage{bm}
\usepackage{amsmath}
\usepackage{autobreak}
\usepackage{hyperref}
\usepackage[mathlines]{lineno}
\usepackage{natbib}
\usepackage{multirow}
\usepackage{booktabs}

\begin{document}

\preprint{APS/123-QED}

\title{Constraints on ultraheavy dark matter from the CDEX-10 experiment at the China Jinping Underground Laboratory}

\author{Y.~F.~Wang}
\affiliation{Key Laboratory of Particle and Radiation Imaging (Ministry of Education) and Department of Engineering Physics, Tsinghua University, Beijing 100084}
\author{L.~T.~Yang}\altaffiliation [Corresponding author: ]{yanglt@mail.tsinghua.edu.cn}
\affiliation{Key Laboratory of Particle and Radiation Imaging (Ministry of Education) and Department of Engineering Physics, Tsinghua University, Beijing 100084}
\author{Q. Yue}\altaffiliation [Corresponding author: ]{yueq@mail.tsinghua.edu.cn}
\affiliation{Key Laboratory of Particle and Radiation Imaging (Ministry of Education) and Department of Engineering Physics, Tsinghua University, Beijing 100084}

\author{K.~J.~Kang}
\affiliation{Key Laboratory of Particle and Radiation Imaging (Ministry of Education) and Department of Engineering Physics, Tsinghua University, Beijing 100084}
\author{Y.~J.~Li}
\affiliation{Key Laboratory of Particle and Radiation Imaging (Ministry of Education) and Department of Engineering Physics, Tsinghua University, Beijing 100084}

\author{H.~P.~An}
\affiliation{Key Laboratory of Particle and Radiation Imaging (Ministry of Education) and Department of Engineering Physics, Tsinghua University, Beijing 100084}
\affiliation{Department of Physics, Tsinghua University, Beijing 100084}

\author{Greeshma~C.}
\altaffiliation{Participating as a member of TEXONO Collaboration}
\affiliation{Institute of Physics, Academia Sinica, Taipei 11529}

\author{J.~P.~Chang}
\affiliation{NUCTECH Company, Beijing 100084}
\author{H.~Chen}
\affiliation{Key Laboratory of Particle and Radiation Imaging (Ministry of Education) and Department of Engineering Physics, Tsinghua University, Beijing 100084}

\author{Y.~H.~Chen}
\affiliation{YaLong River Hydropower Development Company, Chengdu 610051}
\author{J.~P.~Cheng}
\affiliation{Key Laboratory of Particle and Radiation Imaging (Ministry of Education) and Department of Engineering Physics, Tsinghua University, Beijing 100084}
\affiliation{School of Physics and Astronomy, Beijing Normal University, Beijing 100875}
\author{J.~Y.~Cui}
\affiliation{Key Laboratory of Particle and Radiation Imaging (Ministry of Education) and Department of Engineering Physics, Tsinghua University, Beijing 100084}
\author{W.~H.~Dai}
\affiliation{Key Laboratory of Particle and Radiation Imaging (Ministry of Education) and Department of Engineering Physics, Tsinghua University, Beijing 100084}
\author{Z.~Deng}
\affiliation{Key Laboratory of Particle and Radiation Imaging (Ministry of Education) and Department of Engineering Physics, Tsinghua University, Beijing 100084}
\author{Y.~X.~Dong}
\affiliation{Key Laboratory of Particle and Radiation Imaging (Ministry of Education) and Department of Engineering Physics, Tsinghua University, Beijing 100084}
\author{C.~H.~Fang}
\affiliation{College of Physics, Sichuan University, Chengdu 610065}

\author{H.~Gong}
\affiliation{Key Laboratory of Particle and Radiation Imaging (Ministry of Education) and Department of Engineering Physics, Tsinghua University, Beijing 100084}
\author{Q.~J.~Guo}
\affiliation{School of Physics, Peking University, Beijing 100871}
\author{T.~Guo}
\affiliation{Key Laboratory of Particle and Radiation Imaging (Ministry of Education) and Department of Engineering Physics, Tsinghua University, Beijing 100084}
\author{X.~Y.~Guo}
\affiliation{YaLong River Hydropower Development Company, Chengdu 610051}
\author{L.~He}
\affiliation{NUCTECH Company, Beijing 100084}
\author{J.~R.~He}
\affiliation{YaLong River Hydropower Development Company, Chengdu 610051}

\author{H.~X.~Huang}
\affiliation{Department of Nuclear Physics, China Institute of Atomic Energy, Beijing 102413}
\author{T.~C.~Huang}
\affiliation{Sino-French Institute of Nuclear and Technology, Sun Yat-sen University, Zhuhai 519082}

\author{S.~Karmakar}
\altaffiliation{Participating as a member of TEXONO Collaboration}
\affiliation{Institute of Physics, Academia Sinica, Taipei 11529}

\author{Y.~S.~Lan}
\affiliation{Key Laboratory of Particle and Radiation Imaging (Ministry of Education) and Department of Engineering Physics, Tsinghua University, Beijing 100084}
\author{H.~B.~Li}
\altaffiliation{Participating as a member of TEXONO Collaboration}
\affiliation{Institute of Physics, Academia Sinica, Taipei 11529}
\author{H.~Y.~Li}
\affiliation{College of Physics, Sichuan University, Chengdu 610065}
\author{J.~M.~Li}
\affiliation{Key Laboratory of Particle and Radiation Imaging (Ministry of Education) and Department of Engineering Physics, Tsinghua University, Beijing 100084}
\author{J.~Li}
\affiliation{Key Laboratory of Particle and Radiation Imaging (Ministry of Education) and Department of Engineering Physics, Tsinghua University, Beijing 100084}
\author{M.~C.~Li}
\affiliation{YaLong River Hydropower Development Company, Chengdu 610051}
\author{Q.~Y.~Li}
\affiliation{College of Physics, Sichuan University, Chengdu 610065}
\author{R.~M.~J.~Li}
\affiliation{College of Physics, Sichuan University, Chengdu 610065}
\author{X.~Q.~Li}
\affiliation{School of Physics, Nankai University, Tianjin 300071}
\author{Y.~L.~Li}
\affiliation{Key Laboratory of Particle and Radiation Imaging (Ministry of Education) and Department of Engineering Physics, Tsinghua University, Beijing 100084}
\author{Y.~F.~Liang}
\affiliation{Key Laboratory of Particle and Radiation Imaging (Ministry of Education) and Department of Engineering Physics, Tsinghua University, Beijing 100084}

\author{B.~Liao}
\affiliation{School of Physics and Astronomy, Beijing Normal University, Beijing 100875}
\author{F.~K.~Lin}
\altaffiliation{Participating as a member of TEXONO Collaboration}
\affiliation{Institute of Physics, Academia Sinica, Taipei 11529}
\author{S.~T.~Lin}
\affiliation{College of Physics, Sichuan University, Chengdu 610065}
\author{J.~X.~Liu}
\affiliation{Key Laboratory of Particle and Radiation Imaging (Ministry of Education) and Department of Engineering Physics, Tsinghua University, Beijing 100084}
\author{R.~Z.~Liu}
\affiliation{Key Laboratory of Particle and Radiation Imaging (Ministry of Education) and Department of Engineering Physics, Tsinghua University, Beijing 100084}
\author{S.~K.~Liu}
\affiliation{College of Physics, Sichuan University, Chengdu 610065}
\author{Y.~D.~Liu}
\affiliation{School of Physics and Astronomy, Beijing Normal University, Beijing 100875}
\author{Y.~Liu}
\affiliation{College of Physics, Sichuan University, Chengdu 610065}
\author{Y.~Y.~Liu}
\affiliation{School of Physics and Astronomy, Beijing Normal University, Beijing 100875}
\author{H.~Ma}
\affiliation{Key Laboratory of Particle and Radiation Imaging (Ministry of Education) and Department of Engineering Physics, Tsinghua University, Beijing 100084}
\author{Y.~C.~Mao}
\affiliation{School of Physics, Peking University, Beijing 100871}
\author{A.~Mureed}
\affiliation{College of Physics, Sichuan University, Chengdu 610065}
\author{H.~Pan}
\affiliation{NUCTECH Company, Beijing 100084}
\author{N.~C.~Qi}
\affiliation{YaLong River Hydropower Development Company, Chengdu 610051}
\author{J.~Ren}
\affiliation{Department of Nuclear Physics, China Institute of Atomic Energy, Beijing 102413}
\author{X.~C.~Ruan}
\affiliation{Department of Nuclear Physics, China Institute of Atomic Energy, Beijing 102413}
\author{M.~B.~Shen}
\affiliation{YaLong River Hydropower Development Company, Chengdu 610051}
\author{H.~Y.~Shi}
\affiliation{College of Physics, Sichuan University, Chengdu 610065}
\author{M.~K.~Singh}
\altaffiliation{Participating as a member of TEXONO Collaboration}
\affiliation{Institute of Physics, Academia Sinica, Taipei 11529}
\affiliation{Department of Physics, Banaras Hindu University, Varanasi 221005}
\author{T.~X.~Sun}
\affiliation{School of Physics and Astronomy, Beijing Normal University, Beijing 100875}
\author{W.~L.~Sun}
\affiliation{YaLong River Hydropower Development Company, Chengdu 610051}
\author{C.~J.~Tang}
\affiliation{College of Physics, Sichuan University, Chengdu 610065}
\author{Y.~Tian}
\affiliation{Key Laboratory of Particle and Radiation Imaging (Ministry of Education) and Department of Engineering Physics, Tsinghua University, Beijing 100084}
\author{H.~F.~Wan}
\affiliation{Key Laboratory of Particle and Radiation Imaging (Ministry of Education) and Department of Engineering Physics, Tsinghua University, Beijing 100084}
\author{G.~F.~Wang}
\affiliation{School of Physics and Astronomy, Beijing Normal University, Beijing 100875}
\author{J.~Z.~Wang}
\affiliation{Key Laboratory of Particle and Radiation Imaging (Ministry of Education) and Department of Engineering Physics, Tsinghua University, Beijing 100084}
\author{L.~Wang}
\affiliation{School of Physics and Astronomy, Beijing Normal University, Beijing 100875}
\author{Q.~Wang}
\affiliation{College of Physics, Sichuan University, Chengdu 610065}
\author{Q.~Wang}
\affiliation{Key Laboratory of Particle and Radiation Imaging (Ministry of Education) and Department of Engineering Physics, Tsinghua University, Beijing 100084}
\affiliation{Department of Physics, Tsinghua University, Beijing 100084}

\author{Y.~X.~Wang}
\affiliation{School of Physics, Peking University, Beijing 100871}
\author{H.~T.~Wong}
\altaffiliation{Participating as a member of TEXONO Collaboration}
\affiliation{Institute of Physics, Academia Sinica, Taipei 11529}

\author{Y.~C.~Wu}
\affiliation{Key Laboratory of Particle and Radiation Imaging (Ministry of Education) and Department of Engineering Physics, Tsinghua University, Beijing 100084}
\author{H.~Y.~Xing}
\affiliation{College of Physics, Sichuan University, Chengdu 610065}
\author{K.~Z.~Xiong}
\affiliation{YaLong River Hydropower Development Company, Chengdu 610051}
\author{R.~Xu}
\affiliation{Key Laboratory of Particle and Radiation Imaging (Ministry of Education) and Department of Engineering Physics, Tsinghua University, Beijing 100084}
\author{Y.~Xu}
\affiliation{School of Physics, Nankai University, Tianjin 300071}
\author{T.~Xue}
\affiliation{Key Laboratory of Particle and Radiation Imaging (Ministry of Education) and Department of Engineering Physics, Tsinghua University, Beijing 100084}
\author{Y.~L.~Yan}
\affiliation{College of Physics, Sichuan University, Chengdu 610065}
\author{N.~Yi}
\affiliation{Key Laboratory of Particle and Radiation Imaging (Ministry of Education) and Department of Engineering Physics, Tsinghua University, Beijing 100084}
\author{C.~X.~Yu}
\affiliation{School of Physics, Nankai University, Tianjin 300071}
\author{H.~J.~Yu}
\affiliation{NUCTECH Company, Beijing 100084}
\author{X.~Yu}
\affiliation{Key Laboratory of Particle and Radiation Imaging (Ministry of Education) and Department of Engineering Physics, Tsinghua University, Beijing 100084}
\author{M.~Zeng}
\affiliation{Key Laboratory of Particle and Radiation Imaging (Ministry of Education) and Department of Engineering Physics, Tsinghua University, Beijing 100084}
\author{Z.~Zeng}
\affiliation{Key Laboratory of Particle and Radiation Imaging (Ministry of Education) and Department of Engineering Physics, Tsinghua University, Beijing 100084}

\author{F.~S.~Zhang}
\affiliation{School of Physics and Astronomy, Beijing Normal University, Beijing 100875}

\author{P.~Zhang}
\affiliation{Key Laboratory of Particle and Radiation Imaging (Ministry of Education) and Department of Engineering Physics, Tsinghua University, Beijing 100084}

\author{P.~Zhang}
\affiliation{YaLong River Hydropower Development Company, Chengdu 610051}

\author{Z.~Y.~Zhang}
\affiliation{Key Laboratory of Particle and Radiation Imaging (Ministry of Education) and Department of Engineering Physics, Tsinghua University, Beijing 100084}

\author{M.~G.~Zhao}
\affiliation{School of Physics, Nankai University, Tianjin 300071}

\author{J.~F.~Zhou}
\affiliation{YaLong River Hydropower Development Company, Chengdu 610051}
\author{Z.~Y.~Zhou}
\affiliation{Department of Nuclear Physics, China Institute of Atomic Energy, Beijing 102413}
\author{J.~J.~Zhu}
\affiliation{College of Physics, Sichuan University, Chengdu 610065}

\collaboration{CDEX Collaboration}
\noaffiliation

\date{\today}

\begin{abstract}
We report a search for ultraheavy dark matter (UHDM) with the CDEX-10 experiment at the China Jinping Underground Laboratory. Using a Monte Carlo framework that incorporates Earth shielding effects, we simulated UHDM propagation and energy deposition in p-type point-contact germanium detectors. Analysis of 205.4 kg$\cdot$day exposure in the 0.16--4.16 keVee range showed no excess above background. Our results exclude the spin-independent UHDM-nucleon scattering with two cross section scales, with the UHDM mass from $10^6$ to $10^{11}$ GeV, and provide the most stringent constraints with solid-state detectors below $10^8$ GeV.

\end{abstract}
\maketitle

\section{\label{sec1}Introduction}
Numerous cosmological evidence indicates the existence of dark matter (DM)~\cite{pdg2024,bertone_particle_2005}. One of the most popular candidates of DM is weakly interacting massive particles (WIMPs), with mass from GeV to TeV scales~\cite{pdg2024}. Many experiments have conducted direct detection on WIMPs, such as DarkSide~\cite{darkside}, SuperCDMS~\cite{cdmslite}, XENON~\cite{xenon_nT}, LZ~\cite{lux}, PandaX~\cite{pandax}, and CDEX~\cite{cdex0,cdex1,cdex12014,cdex12016,cdex1b2018,cdex1b_am,cdex1bprl,cdex102018}, yet no DM signal has been observed. Consequently, the theoretical framework and experimental investigations are extended to a broader mass range, one of which is heavier DM particles with masses from 10 TeV to the Planck mass~\cite{UHDMwhitebook}. Theories of supersymmetry explain the existence of such ultraheavy particles~\cite{supersymmetry}. At this scale of mass, DM particles can be generated in the early Universe~\cite{UHDM_the1,UHDM_the2}, through certain mechanisms such as freeze-out~\cite{freezeout1,freezeout2} and freeze-in~\cite{freezein1,freezein2}. Composite particles~\cite{composite} and primordial black holes~\cite{pbh} are also candidates of ultraheavy dark matter (UHDM). Although designed for WIMP detection, direct detection experiments are also capable of searching for UHDM. Constrained by the low dark matter density in the Universe, UHDM exhibits a much lower flux than WIMPs, thus requiring an enhanced interaction cross section for observation. The increased cross section facilitates more energy deposition in Earth or atmosphere, known as the Earth shielding effect, which consequently establishes an upper limit of cross section in direct detection experiments~\cite{earthshield,cdms_UHDM}.

Some direct detection experiments have published their research for UHDM, such as DAMA~\cite{dama_UHDM}, CDMS~\cite{cdms_UHDM}, DEAP-3600~\cite{DEAP3600_UHDM}, LZ~\cite{LZ_UHDM}, XENON~\cite{XENON1T_UHDM,Majorana_UHDM}, and Majorana~\cite{Majorana_UHDM}. To optimize UHDM detection sensitivity, simulated signals are compared with experimental data using detector-specific characteristics, generally pulse shape information for liquid scintillator detectors~\cite{DEAP3600_UHDM} and energy for solid-state detectors. For solid-state detectors, the event count in a specific energy region provides constraints on the UHDM mass and cross section~\cite{dama_UHDM,cdms_UHDM,Majorana_UHDM}. The p-type point-contact germanium ($p$PCGe) detector utilized by the CDEX Collaboration is suitable for UHDM research because of its ultralow energy threshold and excellent energy resolution~\cite{darkside}. CDEX-10, the second generation of experiment, consists of a 10 kg $p$PCGe detector array~\cite{cjpl}. Detectors are immersed in liquid nitrogen, with additional shield made up by high-purity oxygen-free copper, lead, and polyethylene~\cite{cdex102018}. The facilities are located in China Jinping Underground Laboratory (CJPL) with a rock overburden of 2400 m (6720 mwe)~\cite{cjpl}. The comprehensive passive shielding system establishes an ultralow radiation background for the detector, about 2.5 $\mathrm{counts \cdot kg^{-1} \cdot keVee^{-1}  \cdot day^{-1}}$ (cpkkd) at an energy range of 0.16--4.16 keV~\cite{cdex10_tech,cdex102020}.

In this work, we focus on the spin-independent elastic scatter between pointlike UHDM and nucleus. We use a Monte Carlo method to simulate the energy deposition of UHDM in the CDEX-10 detector. We systematically analyze the Earth shielding effect's impact on the expected UHDM energy spectrum. Finally, we perform $\chi^{2}$ fitting between the UHDM spectrum and 205.4 kg$\cdot$day dataset from CDEX-10 and establish constraints for UHDM mass and cross section.

\section{\label{sec2}Derivation of UHDM Spectrum}
In order to obtain the expected energy spectrum of UHDM, it is necessary to simulate the entire process of UHDM particles entering the detector. This simulation can generally be divided into three components: the generation of UHDM particles, velocity attenuation during propagation, and energy deposition within the detector. We integrate these three steps into one simulation program and implement each component through Monte Carlo methods.

\subsection{\label{sec2.1}Procedures}
Since the UHDM is much heavier than any atomic nucleus, its velocity direction remains essentially unchanged in the laboratory frame after collisions, resulting in straight-line trajectories. This linear propagation model is applied throughout the simulation, and its scope of applicability will be discussed in Sec.~\ref{sec3.2}. In this model, all simulated UHDM particles capable of detection must have velocity vectors pointing directly toward the laboratory. The velocity distribution of UHDM follows the standard halo model~\cite{SHM}. The value of the local density of UHDM $\rho_{\chi}$ is 0.3 $\mathrm{GeV/c^2/cm^3}$, the escape speed in the Galactic frame is $v_{esc} = 544~\mathrm{km/s}$, and $v_0 = 238~\mathrm{km/s}$ is the local standard of rest at the location of the Sun~\cite{SHM_fv}. The laboratory's mean velocity relative to the Galactic frame arise from the Sun's orbital motion around the Galactic Center, Earth's revolution around the Sun, and the rotation of Earth~\cite{rotation}. This can be calculated by the transformation matrix~\cite{rotation,transmatrix}. A three-dimensional acceptance-rejection algorithm is used to generate the UHDM following the velocity distribution above.

The cross section between UHDM and nucleus ($\sigma_{\chi-N}$) needs discussion before the following simulation. Born approximation gives the model-independent scaling relation for the spin-independent elastic scattering cross section~\cite{born1,born2}:
\begin{equation} \label{A4_sigscale}
  \sigma_{\chi-N} = A^{2}\frac{\mu_N^2}{\mu_n^2}\sigma_{\chi-n},
\end{equation}
where $\sigma_{\chi-n}$ is the cross section between UHDM and nucleon and $\mu$ is the reduced mass for nucleon ($n$) and nucleus ($N$). For UHDM, the relation can be further approximated as scaling with $A^4$. However, the Born approximation breaks down for cross sections exceeding $10^{-25}~\mathrm{cm^2}$~\cite{born1}, and there is currently no unified theoretical framework for scatters at such a cross section. Therefore, two different descriptions of cross section are generally used to compare with other experiments~\cite{DEAP3600_UHDM,Majorana_UHDM}, which are the UHDM per-nucleon scaling ($A^4$ scaling) and the UHDM per-nucleus scaling (A-independent scaling). Separate simulations are conducted for both cross section models.

The second step is calculating the velocity decline from the Earth shielding effect (ESS). When $\sigma_{\chi-N}$ reaches $10^{-26}~\mathrm{cm^2}$ scale, which is the cross section we consider, the mean free path of UHDM particles is approximately 1 m, significantly smaller than Earth's dimensions. Therefore, we adopt a continuous energy attenuation method in ESS. The velocity decline per unit distance is calculated as follows, given by Ref.~\cite{cdms_UHDM}:

\begin{equation} \label{velocity_annu}
   \frac{d v}{d D}=-\frac{m_{p}}{m_{\chi}} v \sum_{i} n_{i}(\boldsymbol{r}) A_{i} \int_{0}^{1} 2 x \sigma_{\chi-N}\left(x E_{i}^{\max }\right) d x,
\end{equation}
where $i$ represents the species at position $\boldsymbol{r}$, $n_i(\boldsymbol{r})$ is the number density, and $\sigma_{\chi-N}(E)$ is the cross section, with differential descriptions mentioned before.

The entire shielding system in the simulation comprises the atmosphere, Earth, the Jinping mountain, and the lead shield outside the detector. Only elements $\rm{N_2}$ and $\rm{O_2}$ are considered for the atmosphere, and the density profile is taken from Ref.~\cite{ISO}. The structures of Earth and the Jinping mountain are adopted from Ref.~\cite{JPmountain}. The density of the mountain is set to 2.7 $\mathrm{g/cm^3}$, while the density and elemental composition of Earth are taken from Ref.~\cite{earth}. The laboratory is located on the surface of Earth, with the Jinping mountain above.  An additional isotropic 20 cm lead shield is placed before particles reach the detector. Copper and polyethylene exhibit negligible shielding effects in our cross section range and are, therefore, not included in the simulation.

After particle generation, its trajectory is treated as a straight line. The simulation step is initially set to 100 m in the atmosphere, Earth and the Jinping mountain. During the simulation, the particle's positions before and after each step are checked. If a change in logical volume is detected (e.g., from atmosphere to mountain), the step length is reduced to one-tenth of its current value, down to a minimum of 0.1 m. The simulation step within the lead shielding is fixed at 5 mm.

And for the last step, we simulate the energy deposition process in a germanium detector. The particle is resampled on a spherical surface enveloping the cylindrical detector to calculate its track lengths within the detector ($l_{\rm{Ge}}$). The mean free path is calculated by
\begin{equation} \label{meanpath}
  \frac{1}{\lambda_{\rm{Ge}}} = n_{\rm{Ge}}\sigma_{\chi-\rm{Ge}}(0)\int_{0}^{1} F^2(xE_R^{\max})dx,
\end{equation}
where $\sigma_{\chi-\rm{Ge}}(0)$ is the cross section with zero momentum transfer. $F^2(E_R)$ is the form factor proposed by Helm~\cite{HELM}. $\bar{N} = \frac{l_{\rm{Ge}}}{\lambda_{\rm{Ge}}}$ represents the average number of collisions in the detector, which reaches $10^3$ times when $\sigma_{\chi-\rm{Ge}}(0) = 10^{-20}~\mathrm{cm^2}$. In the sparse collision regime ($\bar{N}<10^3$), we simulate each collision in the detector, with step and energy deposition given by
\begin{equation} \label{stepGe}
  \rm{step} = -\lambda_{\rm{Ge}}\rm{ln}(1-\xi), \xi \sim U(0,1),
\end{equation}
\begin{equation} \label{Edistribution}
  f(E_R) = \frac{F^2(E_R)}{\int_{0}^{E_R^{\max}}F^2(E_R)d(E_R)}, E_R\in[0,E_R^{\max}],
\end{equation}
\begin{equation} \label{elec_equalized}
  E_v = E_R\cdot f_n(E_R),
\end{equation}
where $f(E_R)$ is the probability for the scatter with recoil energy $E_R$. $E_v$ is the electron-equivalent energy using the quenching factor $f_n(E_R)$ from the Lindhard model~\cite{lindhard}, ensuring consistency with other experimental analyses. Energy below the bandwidth 0.67 eV is rejected. We sum the energy of each step as the energy of one event and incorporate the CDEX-10 energy resolution model for the detector energy response. When the scatter times increase ($\bar{N}>10^3$), the continuous deposition model is used to accelerate the simulation. The expect electron-equivalent energy in one step is given by
\begin{equation}  \label{con_deposition}
\begin{aligned}
  \left\langle E_{v}\right\rangle &=\int_{E_{v}^{\min }}^{E_{v}^{\max }} E_{v} g\left(E_{v}\right) d E_{v}\\
  &=\int_{E_{R}^{\min }}^{E_{R}^{\max }} E_{R} \cdot f_{n}\left(E_{R}\right) f\left(E_{R}\right) d E_{R},
\end{aligned}
\end{equation}
where $g\left(E_{v}\right)$ is the probability for the electron-equivalent energy, which is related to $f(E_R)$ as follows:
\begin{equation} \label{prob_relation}
  g\left(E_{v}\right) = f\left(E_{R}\right)\cdot \left | \frac{dE_{v}}{dE_{R}} \right |.
\end{equation}

\subsection{\label{sec2.2}Results and discussion}
\begin{figure}[!tbp]
\includegraphics[width=\linewidth]{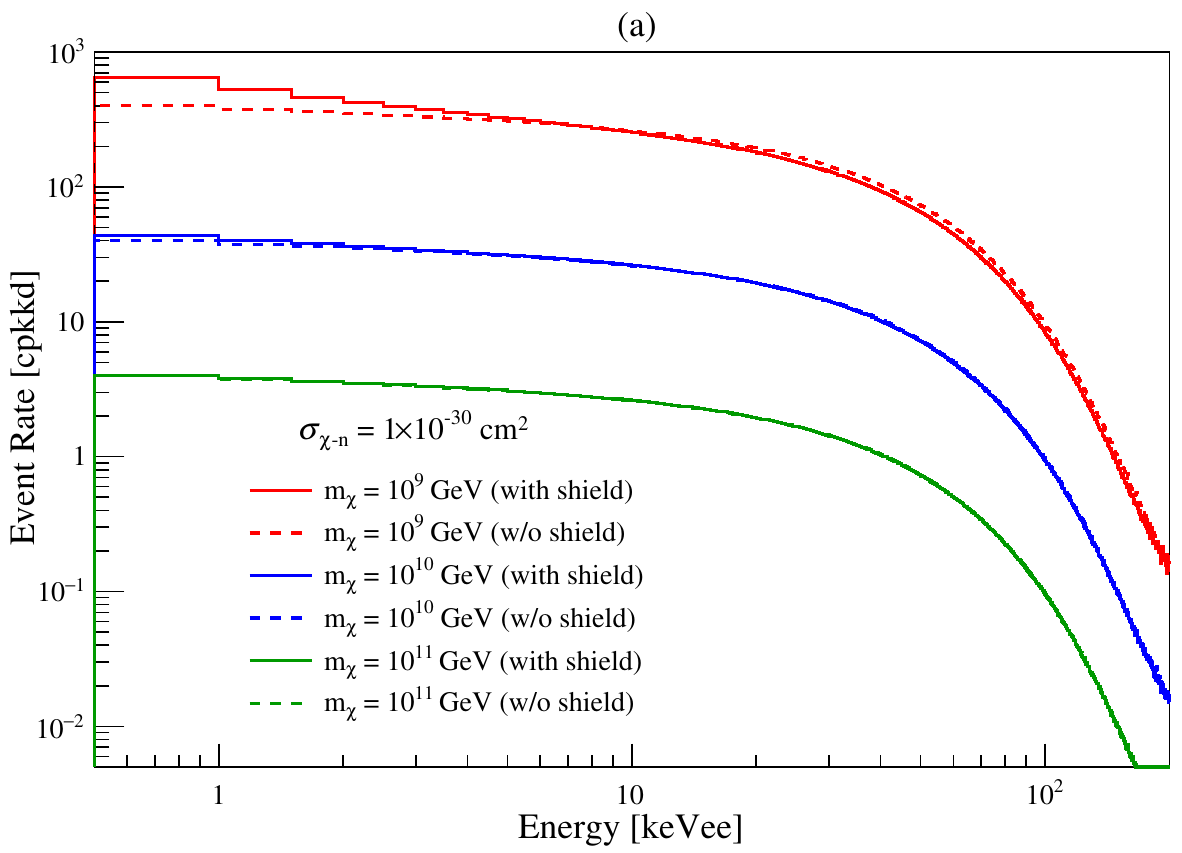}
\includegraphics[width=\linewidth]{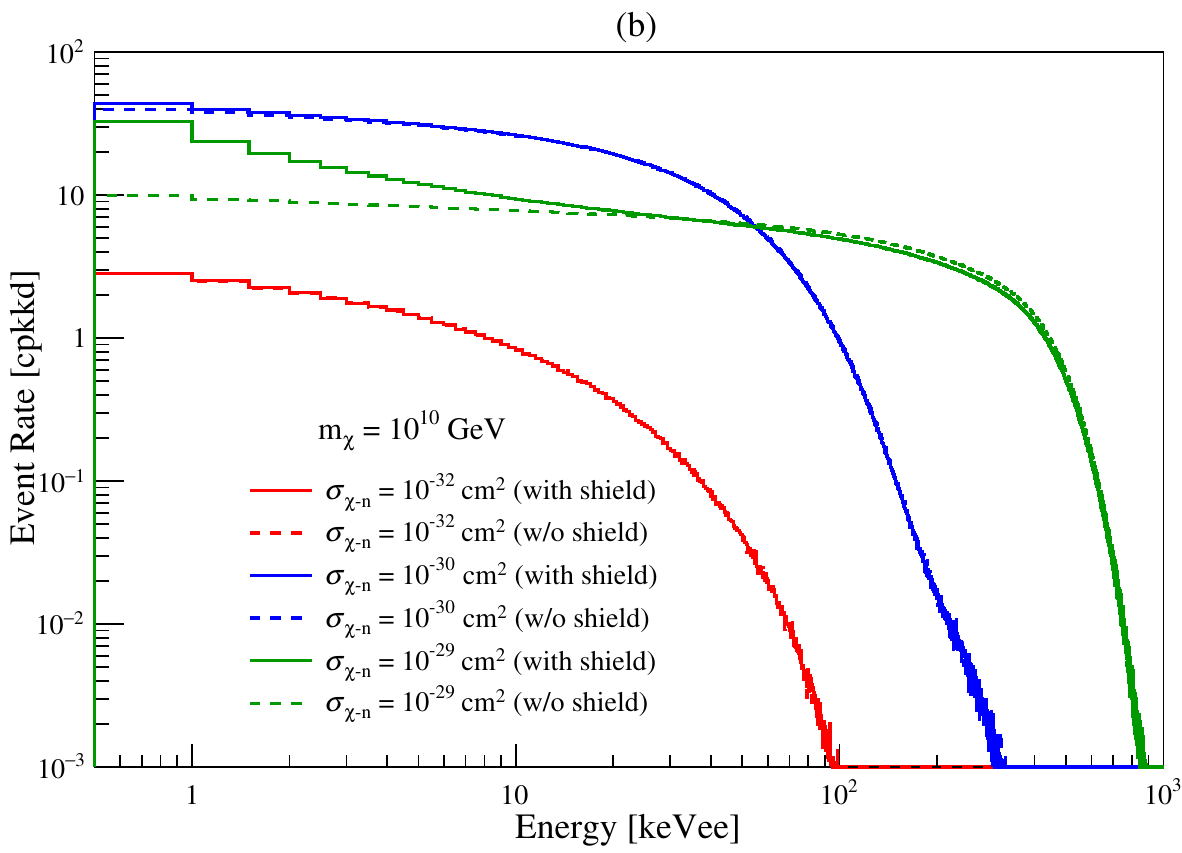}
\caption{Expectation spectrum for UHDM with $A^4$ scale cross section. The parameters are settled for (a) identical cross section as $\sigma_{\chi - n} = 10^{-30}~\mathrm{cm^2}$, and different particle mass including $10^{9}$, $10^{10}$, and $10^{11}~\mathrm{GeV}$, and (b) identical mass as $10^{10}~\mathrm{GeV}$, and different cross section including $10^{-32}$, $10^{-30}$, and $10^{-29}~\mathrm{cm^2}$. Dashed lines represent the spectra without the ESS process. The particles with $\sigma_{\chi-n} = 10^{-32}~\mathrm{cm^2}$ are not affected by ESS, so the red dashed line coincides with the solid line.}
\label{fig::woshield}
\end{figure}

The shape of the expectation spectrum of UHDM under various cross sections merits further examination. Figure~\ref{fig::woshield} shows the spectra with and without the Earth shielding process. Simulation results indicate that, when the ESS is not taken into account and the mass is determined, the expectation spectrum is affected by the cross section with two phases. First, when the cross section is below $10^{-30}~\mathrm{cm^2}$, the probability of UHDM scattering within the detector increases with the cross section, leading to an elevation of event rate. For the second phase, when the cross section exceeds $10^{-30}~\mathrm{cm^2}$, the number of scatters increases with the cross section, leading to a suppression of event rate in the low-energy region, and the spectrum extending to higher energies. The mass of UHDM particles primarily influences the overall event rate of the energy spectrum. Therefore, in the following discussion, we fix the particle mass at $10^{10}~\mathrm{GeV}$. ESS reduces the velocity of UHDM particles reaching the detector, inducing multiple enhancements in counting rate within the low-energy regime, particularly near the detector threshold.

\begin{figure}[!htbp]
\includegraphics[width=\linewidth]{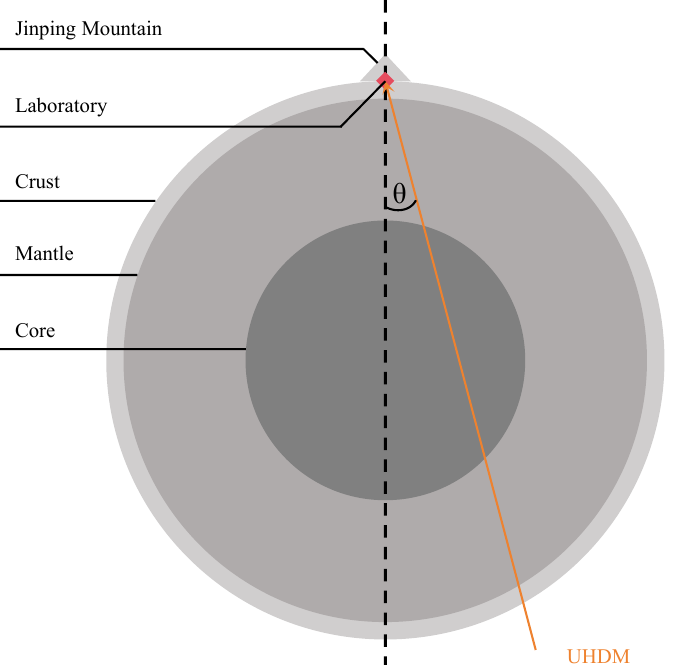}
\caption{
  Structure of Earth and the path of a UHDM particle (yellow array). The structure species it crosses is determined by the angle $\mathrm{\theta}$.
}
\label{fig::earth}
\end{figure}

\begin{figure*}[!htbp]
\includegraphics[width=0.95\linewidth]{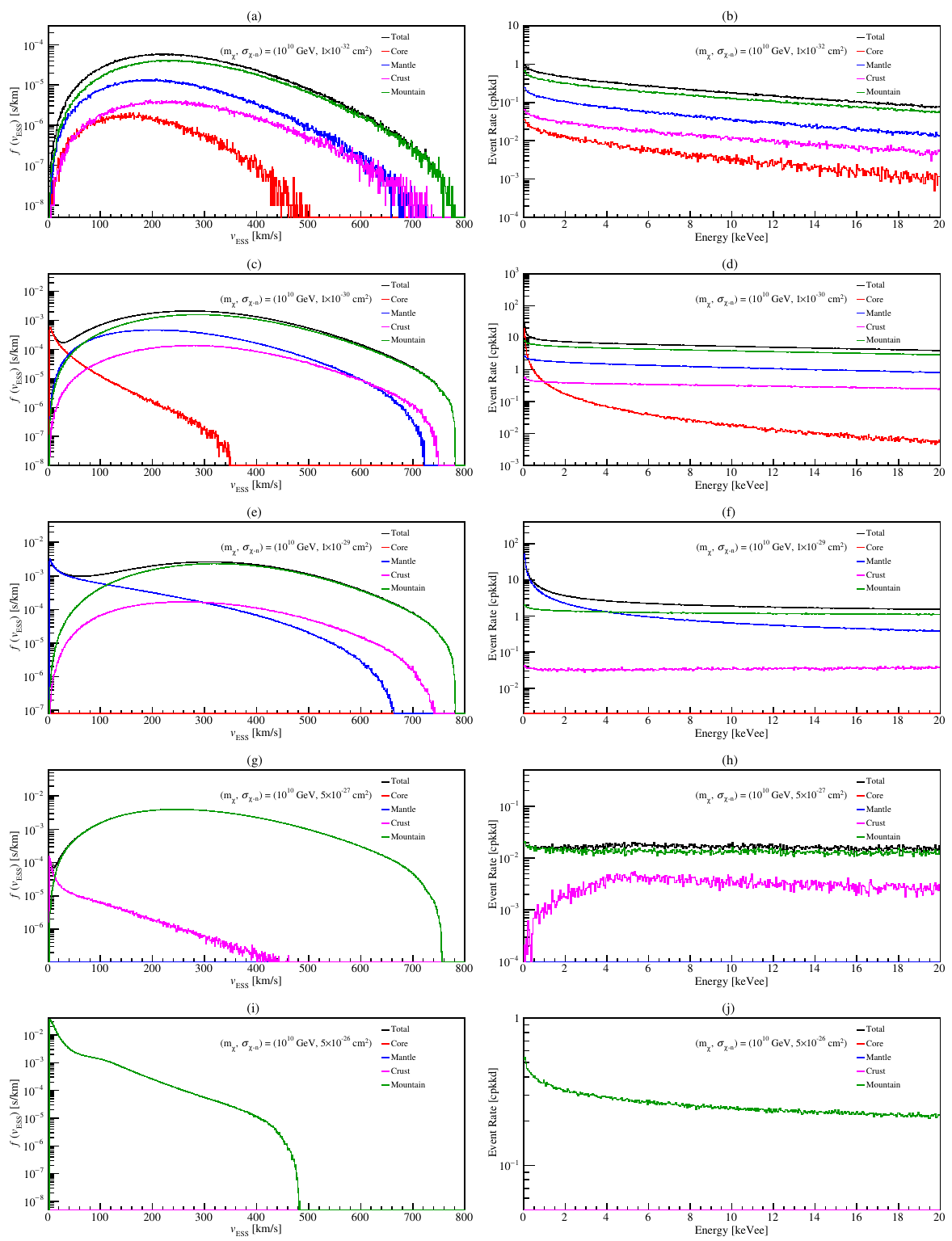}
\caption{The velocity distribution after shielding [$f(v_{\mathrm{ESS}})$] and expectation spectrum of UHDM contributed by different components. The mass of particle is settled as $10^{10}~\mathrm{GeV}$. $A^4$ scale is used here, and cross section $\sigma_{\chi-n}$ is (a),(b) $1\times 10^{-32}$, (c),(d) $1\times 10^{-30}$, (e),(f) $1\times 10^{-29}$, (g),(h) $5\times 10^{-27}$, and (i),(j) $5\times 10^{-26}~\mathrm{cm^2}$. As the cross section increases, UHDM particles of different components become the dominant contributors to the energy spectrum, respectively, while their velocity distributions exhibit an increased probability below 100 km/s, contributed by the ESS.}
\label{fig::spectrum}
\end{figure*}

This phenomenon arises from the structure of Earth's shield. Earth's shielding structure can be categorized into the core, mantle, crust, and mountain, which exhibiting significant density variations. Based on the types of shielding structures traversed by UHDM particles, we classify them into four distinct categories. Because of the linear propagation, these categories can be distinguished by the incident angle of particles. The density of four structures and their angle range are shown in Fig.~\ref{fig::earth} and Table~\ref{tab:table1}. Then we give the distributions of the velocity after attenuated in shielding ($v_{\mathrm{ESS}}$) and energy spectra for each category at specific cross section. Five representative cross sections were selected to elucidate the Earth shielding effect. ESS sequentially affects particles through the core, mantle, crust, and mountain, inducing progressive velocity attenuation and enhancing the counting rate in the region of several keV. Specifically, since the scale of the mountain is significantly smaller than Earth's radius, particles require a much higher cross section to exhibit velocity attenuation within the mountain. As demonstrated in Fig.~\ref{fig::spectrum}, particles through Earth are almost entirely attenuated at a cross section of $5\times 10^{-27}~\mathrm{cm^2}$, whereas particles through the mountain begin to be affected when the cross section reaches $5\times 10^{-26}~\mathrm{cm^2}$.

\begin{table}[!htbp]
\caption{\label{tab:table1}Categories of UHDM and their angle ranges.}
\renewcommand\arraystretch{1.5}
\begin{ruledtabular}
\begin{tabular}{lcc}
  Category & Shielding components               & $\mathrm{\theta}$ range /rad \\
\hline
  Core	   & Atmosphere, core, mantle, crust, lead	  & [0, 0.575] \\
  Mantle	 & Atmosphere, mantle, crust, lead	        & [0.575, 1.441] \\
  Crust	   & Atmosphere, crust, lead	                & [1.441, 1.573] \\
  Mountain & Atmosphere, mountain, lead	              & [1.573, 3.141] \\
\end{tabular}
\end{ruledtabular}
\end{table}

\begin{figure}[!htbp]
\includegraphics[width=\linewidth]{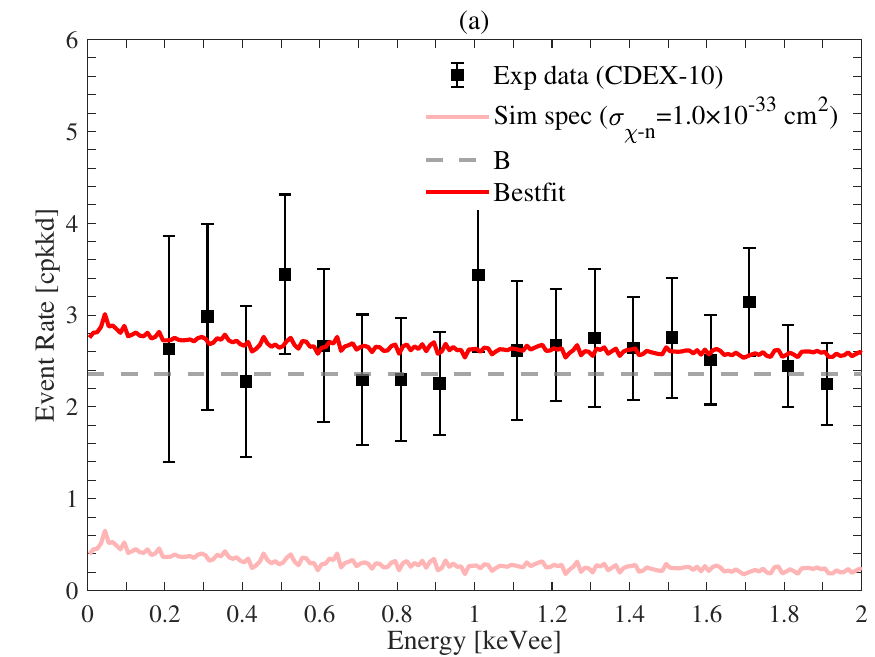}
\includegraphics[width=\linewidth]{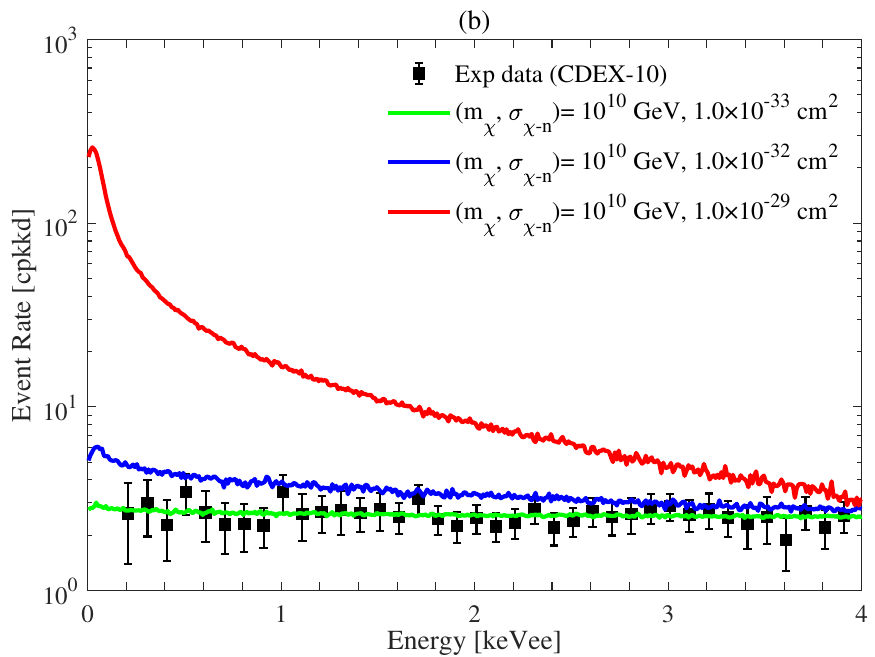}
\caption{(a) The best-fit result for the simulation spectrum with the parameters of $m_{\chi} = 10^{10}~\mathrm{GeV}$ and $\sigma_{\chi - n} = 10^{-33}~\mathrm{cm^2}$, using the $A^4$ cross section scale. The CDEX-10 experimental residual spectrum is used in this fit. The energy resolution is considered in this plot, with the standard deviation settled as $\sigma = 35.8 + 16.6\times \sqrt{E}~ (\mathrm{eV})$~\cite{XR}, where $E$ is in keV.
(b) Best-fit result for the simulation spectra with same mass ($m_{\chi} = 10^{10}~\mathrm{GeV}$) and different cross section. By using the minimum-$\chi^2$, parameter $\sigma_{\chi - n} = 10^{-33}~\mathrm{cm^2}$ is under the exclusion area, while $\sigma_{\chi - n} = 10^{-32}~\mathrm{cm^2}$ and $\sigma_{\chi - n} = 10^{-30}~\mathrm{cm^2}$ are within the $90\%$ CL exclusion area.}
\label{fig::specfit}
\end{figure}

\section{\label{sec3}Data analysis}
\subsection{\label{sec3.1}Procedures}
In this work, we use the CDEX-10 data taken from February 2017 to August 2018, with exposure of 205.4 kg$\cdot$day~\cite{cdex102020}. The data processing procedure follows the methodology established in the Collaboration's previous publications, including energy calibration, physical event selection, bulk-surface event discrimination, and efficiency corrections~\cite{cdex102018,cdex10_tech,cdex102020}. At 0--12 keVee energy region, the radiation background is mainly consist of Compton scattering of high-energy gamma rays and internal radioactivity from long-lived cosmogenic isotopes~\cite{ZZY}. Our Collaboration established a background model at such region, fitting the K-shell characteristic x-ray peaks from the internal cosmogenic radionuclides, and deriving the L-shell x-ray peaks through their corresponding K-shell line intensities~\cite{kshell,XR}. Background after subtracting the L- and K-shell peaks are utilized for the UHDM physics analysis, as shown in Fig.~\ref{fig::specfit}. A minimum-$\chi^2$ method~\cite{cdex12014} is applied to the residual spectrum at the range of 0.16--4.16 keVee. The $\chi^2$ statistic was constructed in the form of Eq.~\ref{chi2}, incorporating the UHDM expected energy spectrum $S_i$ and the flat background $B$~\cite{ZZY}:

\begin{equation} \label{chi2}
  \chi^{2}\left(m_{\chi}, \sigma_{\chi-N}\right)=\sum_{i=1}^{N} \frac{\left[N_{i}-B-S_{i}\left(m_{\chi}, \sigma_{\chi-N}\right)\right]^{2}}{\Delta_{i}^{2}}.
\end{equation}

For specific parameter $\left ( m_{\chi},\sigma _{\chi-N}  \right ) $, a positive best fit $B$ is calculated for the minimum of $\chi^2\left ( m_{\chi},\sigma _{\chi-N}  \right ) $. For the whole parameter space, a $90\%$ confidence level (CL) one-side upper limit exclusion area with $\Delta \chi^2 = 1.64$ is derived. 

\begin{figure}[!htbp]
\includegraphics[width=\linewidth]{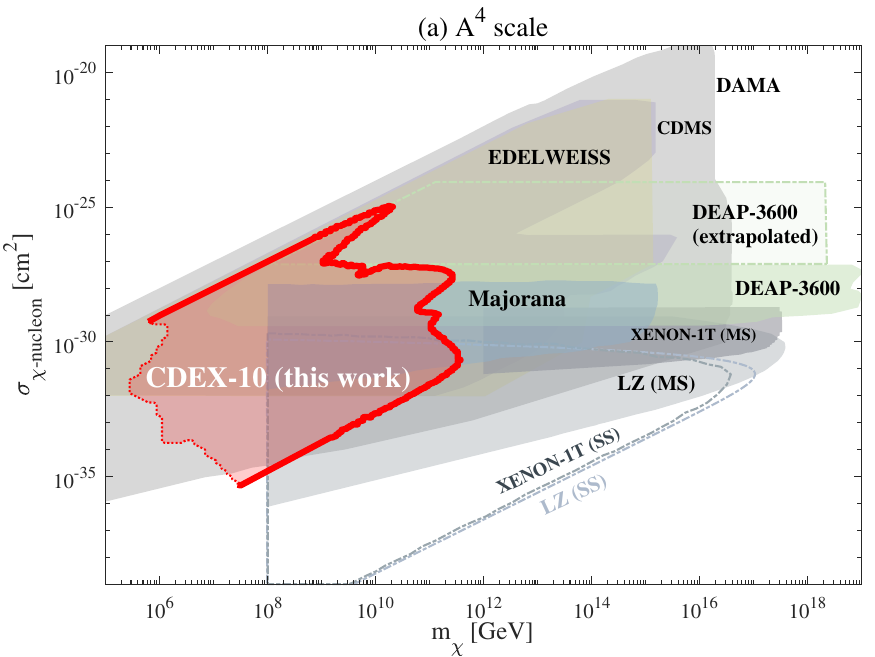}
\includegraphics[width=\linewidth]{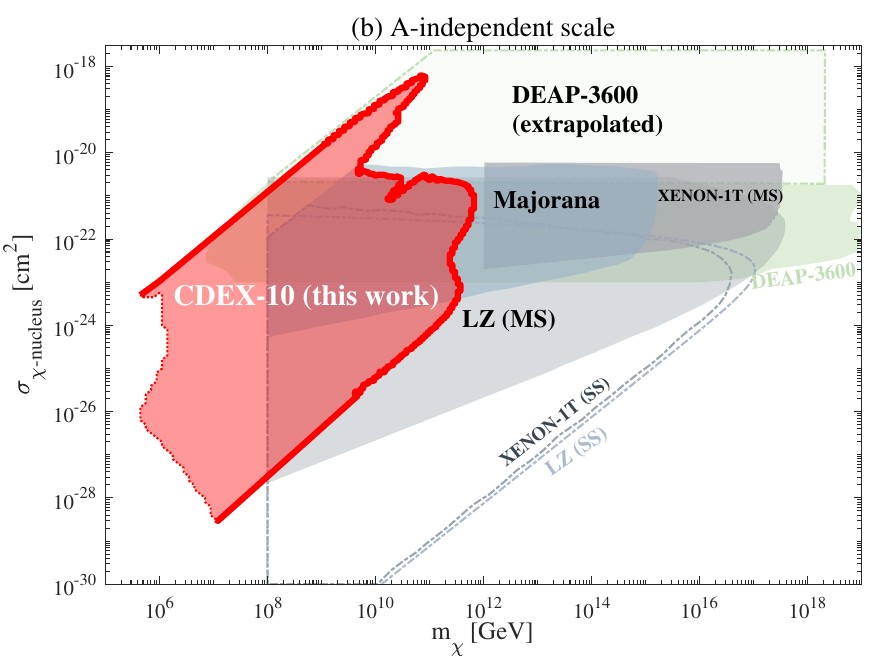}
\caption{Exclusion regions with (a) the $A^4$ scale for UHDM mass vs UHDM-nucleon cross section and (b) the $A$-independent scale for UHDM mass vs UHDM-nucleus cross section. A $90\%$ confidence level (CL) one-side upper limit is provided for both exclusions. Our results are compared with those from other direct detection experiments, including Majorana~\cite{Majorana_UHDM}, DAMA~\cite{dama_UHDM}, CDMS and EDELWEISS~\cite{cdms_UHDM,cdms0_UHDM} using solid-state detectors, as well as XENON-1T~\cite{Majorana_UHDM,XENON1T_UHDM}, LZ~\cite{LZ_UHDM}, and DEAP-3600~\cite{DEAP3600_UHDM} using liquid scintillator detectors. XENON and LZ published their multiple-scatter and single-scatter results separately, marked as MS and SS, respectively, in the figure.}
\label{fig::exclusion}
\end{figure}

\subsection{\label{sec3.2}Constraints and discussion}
Figure~\Ref{fig::exclusion} shows the exclusion regions with two scales of cross section, comparing with constraints from other direct detection experiments~\cite{dama_UHDM,cdms_UHDM,DEAP3600_UHDM,Majorana_UHDM,cdms0_UHDM}. Because of Earth's shielding effects, the exclusion upper limits is related to the overburden depth. The shape of the exclusion area at high masses arises from the structure of Earth, discussed in Sec.~\ref{sec2.2}. Compared with other experiments, we present the best exclusion limits below $10^8~\mathrm{GeV}$ for solid-state detectors, achieved through spectral fitting. Specifically, we compare our analysis methodology and results with the Majorana Collaboration, as the similar detection approach of a geranium detector. The Majorana experiment analyzed physical events interacting with multiple detectors in its detector arrays and simulated corresponding UHDM signals, thereby establishing exclusion limits for DM mass above $10^{12}~\mathrm{GeV}$~\cite{Majorana_UHDM}. We analyze the deposition energy at keV scale and present a better lower limit with mass below $10^{11}~\mathrm{GeV}$.

The scope of applicability of the UHDM linear propagation model in Earth shielding is shown as the dashed curve in Fig.~\ref{fig::exclusion} and requires further discussion. When a UHDM particle undergoes single elastic scattering with an atomic nucleus, the relationship between the scattering angles in the center-of-mass frame ($\alpha_{c}$) and the laboratory frame ($\alpha$) is given by
\begin{equation} \label{scattering_angle_c}
  \cos \alpha_{c} = 1 - 2 \frac{E_R}{E_R^{\max}},
\end{equation}
\begin{equation} \label{scattering_angle}
  \tan \alpha = \frac{\sin \alpha_{c}}{\frac{m_{\chi}}{m_{N}}+\cos \alpha_{c}} \simeq \frac{m_N}{m_{\chi}}\sin \alpha_{c},
\end{equation}
where $E_R$ is the nuclear recoil energy, with its distribution given in Eq.~\ref{Edistribution}. This approximation is valid when $m_{\chi}\textgreater10^5~\mathrm{GeV}$. The expected value and variance of $\alpha$ are calculated as
\begin{equation} \label{expectation_alpha}
  \begin{aligned}
  \left\langle\alpha\right\rangle &= \frac{m_N}{m_{\chi}}\frac{\int_{0}^{1} 2\sqrt{x(1-x)} F^2(xE_R^{\max}) dx}{\int_{0}^{1} F^2(xE_R^{\max}) dx}\\
  &= \frac{m_N}{m_{\chi}} \frac{\Theta(m_{\chi},v)}{\Gamma(m_{\chi},v)},
  \end{aligned}
\end{equation}
\begin{equation} \label{variance_alpha}
  \begin{aligned}
  \sigma^2(\alpha) &= \left( \frac{m_N}{m_{\chi}} \right)^2 \frac{\int_{0}^{1} 4x(1-x) F^2(xE_R^{\max}) dx}{\int_{0}^{1} F^2(xE_R^{\max}) dx} - \left\langle\alpha\right\rangle^2\\
  &=\left( \frac{m_N}{m_{\chi}} \right)^2 \left[ \frac{\Delta(m_{\chi},v)}{\Gamma(m_{\chi},v)} - \left( \frac{\Theta(m_{\chi},v)}{\Gamma(m_{\chi},v)} \right)^2 \right].
  \end{aligned}
\end{equation}

The coherence factors $\Theta(m_{\chi},v)$, $\Gamma(m_{\chi},v)$, and $\Delta(m_{\chi},v)$ reflect the influence of the form factor. For pointlike particles, $F^2(q) \to 1$ and $\frac{\Theta(m_{\chi}, v)}{\Gamma(m_{\chi}, v)} \to \pi/4$, which corresponds to isotropic scattering in the center-of-mass frame. By sampling the scattering angle $\alpha$ within [0, $m_A/m_{\chi}$] and the azimuthal angle $\phi$ within [0, $2\pi$], we can simulate the scattering direction at each step. Note that the scattering angle $\alpha$ is very small, so the deviation of the UHDM in the forward direction is negligible. After multiple collisions (typically, $N_{\mathrm{earth}} > 10^4$), the orthogonal components of the radial deviation are given by
\begin{equation} \label{delta_x}
  \begin{aligned}
  \Delta x = \sum_{j=1}^{N_{\rm{earth}}} l_j \sin \alpha_j \cos \phi_j,\\
  \Delta y = \sum_{j=1}^{N_{\rm{earth}}} l_j \sin \alpha_j \sin \phi_j,
  \end{aligned}
\end{equation}
where $l_j$ is the step length for each collision. The distribution of collision distance and scattering angle for each interaction is determined by the nuclide species and number density at the current position. 

Estimation of the radial deviation in a uniform medium is useful for determining simulation parameters and interpreting the simulation results. When a UHDM particle undergoes $\bar{N}$ collisions in the medium, with a similar propagation distance for each step, the variables $(\Delta x,~\Delta y)$ follow a two-dimensional Gaussian distribution according to the central limit theorem, with the variance given by
\begin{equation} \label{expectation_delta}
  \sigma^2(\Delta x) = \sigma^2(\Delta y) = \left( \frac{m_N}{m_{\chi}} \right)^2 \frac{L}{2\Sigma_{\chi-N}(0)} \frac{\Delta(m_{\chi},v)}{\Gamma(m_{\chi},v)^2},
\end{equation}
where $\Sigma_{\chi-N}(0)$ is the macroscopic cross section with zero momentum transfer and $L$ is the total path length. In this case, the radial deviation follows a Rayleigh distribution, with an identical scale parameter. For each combination of the mass and cross section parameters, we simulate the propagation of $10^5$ particles, adopting the velocity that maximizes the radial deviation. The deviation of each particle is constrained by the detector size, indicating that the probability of UHDM particles failing to reach the detector due to transverse deviation is $\mathcal{O}(10^{-5}) $. Thus, a conservative scope of applicability is derived, which defines the left-side boundary in our analysis. Earth shielding effects can limit the maximum propagation distance of UHDM particles within Earth and, thus, affect the shape of the boundary. The experimental data still have the ability to exclude lighter dark matter candidates, but this scenario will not be further analyzed in this manuscript.

\section{\label{sec4}Conclusion}
In this work, we developed a comprehensive Monte Carlo simulation framework to investigate the energy deposition processes of UHDM in $p$PCGe detectors. A more detailed Earth shielding model was adopted, and the effects of different terrestrial structures on dark matter velocity attenuation and the expected energy spectrum were thoroughly analyzed. We also provide a more detailed discussion on the applicable scope of the UHDM linear propagation model. Through the analysis of CDEX-10 experimental data, the first exclusion area for UHDM was established by the CDEX Collaboration, with the best limits for solid-state detectors in direct detection experiments being achieved when $m_{\chi} < 10^8~\mathrm{GeV}$. This result demonstrates that detectors with low background and low energy thresholds exhibit better constraints in UHDM analyses.

CDEX-50, the next generation of the CDEX experiment, is currently in preparation~\cite{CDEX50}. An array of 50 1-kg $p$PCGe detectors units will be operated in liquid nitrogen cryostats at CJPL-II laboratory. The projected background level is expected to reach $10^{-2}~\mathrm{cpkkd}$ at 0--12 keVee, about 2 orders of magnitude lower than the CDEX-10 experiment. We expect significantly strong constraints for lower cross section limits. Analysis of multiple interactions in different detectors will be applied to extend the exclusion area to heavier mass regions.

\acknowledgments
This work was supported by the National Key Research and Development Program of China (Grants No. 2022YFA1605000 and No. 2023YFA1607101) and the National Natural Science Foundation of China (Grants No. 12425507, No. 12322511, No. 12175112, and No. 123B2087). We acknowledge the Center of High Performance Computing, Tsinghua University, for providing the facility support. We thank CJPL and its staff for hosting and supporting the CDEX project. CJPL is jointly operated by Tsinghua University and Yalong River Hydropower Development Company.

\bibliography{UHDM}

\end{document}